# Recursive Tangential-Angular Operator as Analyzer of Synchronized Chaos

A. V. Makarenko

Constructive Cybernetics Research Group, Moscow, Russia

e-mail: avm.science@mail.ru

**Abstract**. A method for the quantitative analysis of the degree and parameters of synchronization of the chaotic oscillations in two coupled oscillators is proposed, which makes it possible to reveal a change in the structure of attractors. The proposed method is tested on a model system of two unidirectionally coupled logistic maps. It is shown that the method is robust with respect to both the presence of a low-intensity noise and a nonlinear distortion of the analyzed signal. Specific features of a rearranged structure of the attractor of a driven subsystem in the example under consideration have been studied.

The problem of detection and analysis of the synchronization of chaotic signals and systems is important both for the theory of dynamical systems and various applications [1]. It should be noted that the term "synchronized chaos" refers to a number of physical phenomena [1-4]. Among these, a special place is occupied by the synchronization of chaotic oscillations, which one of the most extensively studied phenomena of nonlinear dynamics [1, 2, 4]. Analysis of this phenomenon always poses a very interesting problem of evaluating the degree of synchronization of the chaotic oscillations of two coupled oscillators.

A correct solution of this problem requires using an adequate quantitative measure of the degree of synchronization, which must obey several conditions [4], in particular, to have a clear physical meaning and to be universal and robust. Proceeding from the character of the problem of evaluating the degree of chaos synchronization, it would be natural to suggest that the synchronism influences to some extent the structure of oscillations of the coupled oscillators.

Previously, the author proposed an original approach [5, 6] to an analysis of the structure of dynamic processes in the tome domain, which was based on the use of differential-geometric transformations (maps). These maps are generated by a recursive tangential-angular operator $\mathrm{TG}_k^a$, which forms a set of angular functions defined as follows [5]:

$$\mathrm{TG}_k^a = \tan \prod_{i=1}^{k} \arctan \left[ (c_s)_i \frac{d}{dt} \right], \quad \alpha_x^T = c_{ss}\dot{x}, \quad \alpha_y^T = c_{ss}\dot{y}, \quad \varphi_{0,x}^T = \frac{c_{as}c_{ss}\ddot{x}}{1 + c_{ss}^2[\alpha_x^T]^2}, \quad \varphi_{0,y}^T = \frac{c_{as}c_{ss}\ddot{y}}{1 + c_{ss}^2[\alpha_y^T]^2},$$

where $k = 1, \ldots, N_G$ is the order of the angular operator, $c_{\circ s} > 0$ is the scaling coefficient, and the upper dot denotes the differentiation with respect to time $t$. Operator $\mathrm{TG}_k^a$ acts on the dynamic processes $x(t)$ and $y(t)$, which are signals of the driving (master) and driven (slave) subsystems, respectively. The most important quantities for an analysis of the dynamic process structure are $\alpha_\circ^T$ and $\varphi_{0,\circ}^T$. The former value in fact describes the rate properties (instantaneous slope), while the latter value characterizes the nonlinear properties (instantaneous curvature) of the front of the $x(t)$ or $y(t)$ process. The initial processes (prior to the action of operator $\mathrm{TG}_k^a$ and the subsequent calculations) are centered and normalized as follows:

$$x(t) \to \frac{x(t) - \mathrm{M}[x]}{\int_T |x(t)| \mathrm{d}t}, \quad y(t) \to \frac{y(t) - \mathrm{M}[y]}{\int_T |y(t)| \mathrm{d}t}. \tag{1}$$

where $\mathrm{M}[\circ]$ is the mathematical expectation operator and $T$ is the evaluation period. In addition, the functions $f_s = x + y$, $f_r = x - y$ and the analogous quantities $\alpha_s^T$, $\alpha_r^T$ and $\varphi_{0,s}^T$, $\varphi_{0,r}^T$ are introduced and the measures ($L_1$) of their total intensity on $f_\circ$, $\alpha_\circ^T$ and $\varphi_{0,\circ}^T$ are defined as follows [7]:

$$I_\circ^{FT} = \int_T |f_\circ(t)| \mathrm{d}t, \quad I_\circ^{AT} = \int_T |\alpha_\circ^T(t)| \mathrm{d}t, \quad I_\circ^{\Phi T} = \int_T |\varphi_{0,\circ}^T(t)| \mathrm{d}t. \tag{2}$$

Then, the degree of synchronization $\delta$ ($\delta_\circ \in [0,1]$) of processes $x(t)$ and $y(t)$ can be determined using the following measures:

$$\delta_f = \frac{|I_s^{FT} - I_r^{FT}|}{\int_T |x(t)| + |y(t)| \mathrm{d}t}, \quad \delta_\alpha = \frac{|I_s^{AT} - I_r^{AT}|}{\int_T |\alpha_x^T(t)| + |\alpha_y^T(t)| \mathrm{d}t}, \quad \delta_\varphi = \frac{|I_s^{\Phi T} - I_r^{\Phi T}|}{\int_T |\varphi_{0,x}^T(t)| + |\varphi_{0,y}^T(t)| \mathrm{d}t}. \tag{3}$$

Note that $\delta_\circ$ in fact characterizes the symmetry of components of the $[x(t), y(t)]^\mathrm{T}$ vector with respect to their cyclic permutation [8], where symbol $\circ^\mathrm{T}$ denotes the transposition operator. The symmetry is measured in the $\mathrm{P}_0$ space [5, 6].

The proposed approach to evaluation of the chaotic synchronization was applied to investigation of the process of escape from the regime of complete chaos synchronization in a system of two unidirectionally coupled logistic maps:

$$x_{k+1} = 4\lambda x_k(1 - x_k), \quad y_{k+1} = 4\lambda [y_k + \gamma(x_k - y_k)](1 - [y_k + \gamma(x_k - y_k)]), \tag{4}$$

where $x_k$, $y_k$ are the variables of state of the driving and driven process, respectively ($x, y \in [0,1]$); $\gamma$ is the coupling parameter ($\gamma \in [0,1]$); and $\lambda$ is the control parameter ($\lambda \in [0,1]$). The logistic mapping is well known [9] and used as a test model object in nonlinear and chaotic dynamics [3]. The estimations of $\delta_\circ$ and $I_y^{\circ T}$ values were calculated on the interval of $k \in [1 \times 10^5, 3 \times 10^5]$. This shift from $k = 1$ is related to the necessity of minimizing the parasitic influence of a transient process. In addition, all estimations of the analyzed values were averaged over 300 variants of initial conditions $x_1 = \xi_1$, $y_1 = \xi_2$, where $\xi_1, \xi_2 \in (0,1)$ are uncorrelated uniformly distributed random values. This averaging ensured neutralization of the memory effect induced by the initial conditions on the trajectories of processes $x$ and $y$. The coupling parameter was varied in the interval $\gamma \in [0, 0.5]$ at a discretization step of $1 \times 10^{-4}$. The control parameter was set at $\lambda = 0.95$, which corresponded to a regime of developed chaos in system (4) [3]. This choice was explained by the wish to make possible a mutual analysis and ensure the consistency of results obtained in this study and those reported by Shabunin et al. [4].

The robustness of the proposed measure (3) of chaos synchronization was checked by additive mixing of a small noise to the driven subsystem:

$$x_k \to x_k + \eta(\xi_k' - x_k), \quad \eta = 10^{-5}, \quad \xi' \in [0,1], \tag{5}$$

where $\xi'$ is a uniformly distributed random value. The noise effect was determined by observation of a nonmodified sequence of $x$. In addition, the measure (3) of chaos synchronization was also tested for its stability with respect to nonlinear distortion of the signals studied. For this purpose, the observed sequence of the driven subsystem was modified as follows:

$$x_k \to x_k + \sigma_i x_k^2, \quad \sigma_1 = 0.05, \quad \sigma_2 = 0.2. \tag{6}$$

Figure 1 shows the results of calculations of the $\delta_\varphi$ value as a function of the coupling parameter. An analysis of these data shows that the proposed measure actually reveals the escape of system (4) from the regime of complete synchronization. Indeed, $\delta_\varphi$ begins to decrease from unity at $\gamma_c = 0.35$ to $\delta_\phi \approx 0.0272$ at $\gamma_d \approx 0.1051$. Then, $\delta_\varphi$ exhibits a local increase to $\sim 0.0527$ at $\gamma_u \approx 0.0639$, which is followed by a decrease to $\delta_\phi \approx 5 \times 10^{-3}$ at $\gamma_b \approx 0.01$ and this level is approximately retained during the subsequent decrease in the coupling coefficient to zero.

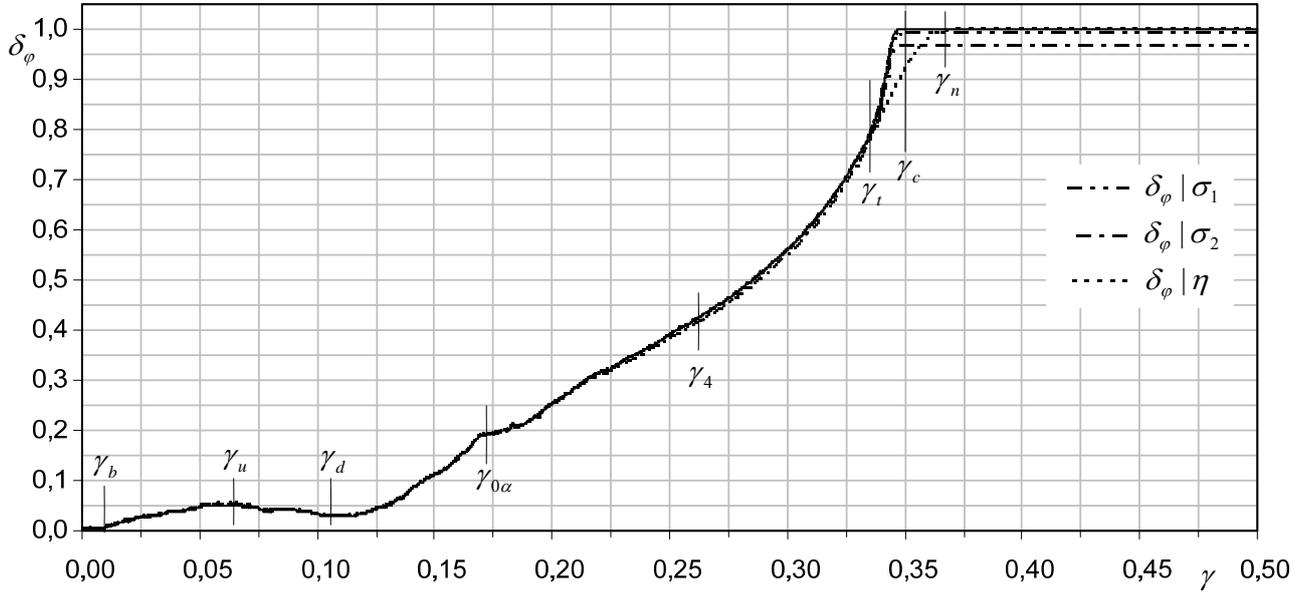

**Fig. 1.** Plot of the synchronization coefficient $\delta_\varphi$ versus coupling parameter $\gamma$.

When a small noise (5) is introduced into the system, the escape from the regime of complete synchronization began earlier ($\gamma_n \approx 0.3684$), which was related to the onset of "bubbling" regime in the driven subsystem [4, 10]. A significant difference between the synchronization coefficients $\delta_\varphi$ of the initial and noisy systems is only observed in the region of $\gamma \in [\gamma_t, \gamma_n]$ (where $\gamma_t \approx 0.3365$), which is related to a nonrobust regime of synchronization in this region [4] (see Fig. 2). Thus, in the general case, the proposed measure $\delta_\varphi$ is stable with respect to the introduction of small noise into the system under consideration.

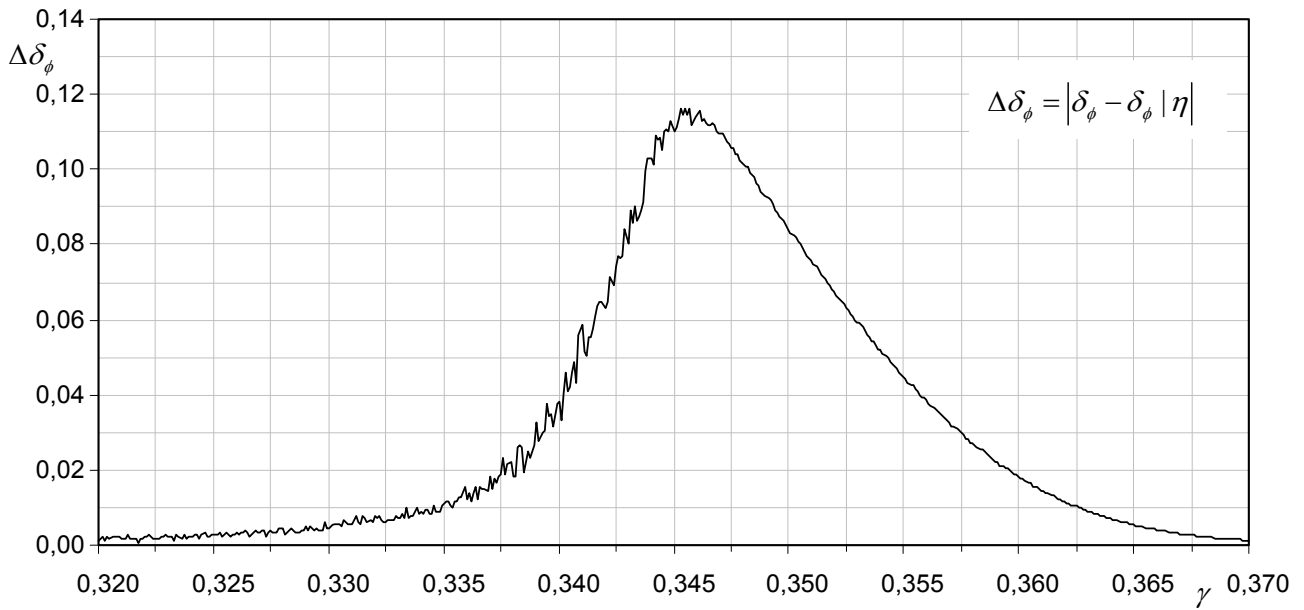

**Fig. 2.** Plot of the difference between values $\delta_\varphi$ and $\delta_\phi | \eta$ versus coupling parameter $\gamma$.

In the case of a nonlinear distortion (6) of the signal, the modified system exhibits a lower final synchronization level at $\gamma > \gamma_c = 0.35$. This result is quite reasonable since, in the case of a generalized synchronization of $y(t) = \mathrm{F}[x(t)]$ with an unknown operator $\mathrm{F}[\circ]$, it is impossible to unambiguously judge between the possible reasons for a decrease in the synchronization coefficient (nonlinear distortion of observables versus incomplete synchronism) [1]. Thus, the proposed measure (3) is also stable with respect to a nonlinear distortion of type (6) and, hence, can be used to study the process of generalized synchronization. Analogous conclusions are valid for the values of $\delta_f$ and $\delta_\alpha$.

Although the measurement of a synchronization coefficient is important, it is by no means sufficient for the adequate description of processes on the system under consideration. It is also necessary to study the structure of attractors formed in the driven subsystem y at various values of the coupling parameter $\gamma$. For example, it was reported [4] that, at $\gamma = 0.14$, the driven subsystem in (4) exhibits a reconstruction of the attractor. Using the proposed approach, it is possible to predict changes in the attractor structure caused by variations in the degree of synchronization in the system. For this purpose, it is necessary to study the behavior of the centered values of intensities $\bar{I}_y^{\circ T}$ as functions of the coupling parameter $\gamma$. This dependence is presented in Fig. 3 and Fig. 4, where some special values of $\gamma$ can be recognized, in particular, $\gamma_1 = 0.1241$, $\gamma_2 = 0.14$, $\gamma_3 = 0.1558$ and $\gamma_4 = 0.2606$. Note that at $\gamma_1$-$\gamma_3$ we observe $I_y^{\circ T} > I_x^{\circ T}$, whereas at $\gamma_4$ we have $I_y^{\circ T} < I_x^{\circ T}$. It should also be recalled that values are calculated using the normalized time series (1) of $x(t)$ and $y(t)$. Therefore, in view of the physical meaning of values, their variations cannot be assigned entirely to the contraction/expansion of the attractor of the driven subsystem (although it is undoubtedly present). Thus, the plots in Fig. 3 and Fig. 4 primarily reflect the structural rearrangement of the attractor of the driven subsystem as a result of a change in the degree of action of (coupling with) the driving subsystem. The value of $\gamma_{0\alpha} = 0.1718$ was considered as a control point (where $I_y^{AT} = I_x^{AT}$), relative to which the observed changes in the structure of attractors at $\gamma_3$ and $\gamma_4$ were analyzed.

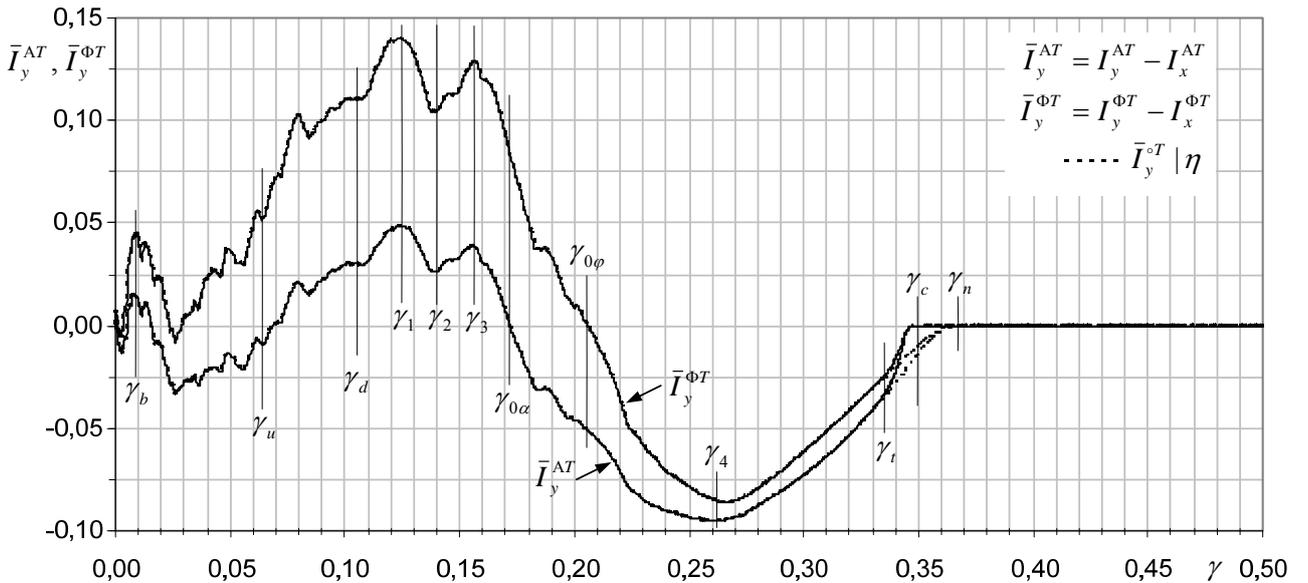

**Fig. 3.** Plot of the intensities $\bar{I}_y^{AT}$ and $\bar{I}_y^{\Phi T}$ versus coupling parameter $\gamma$.
Without using the normalized time series (1).

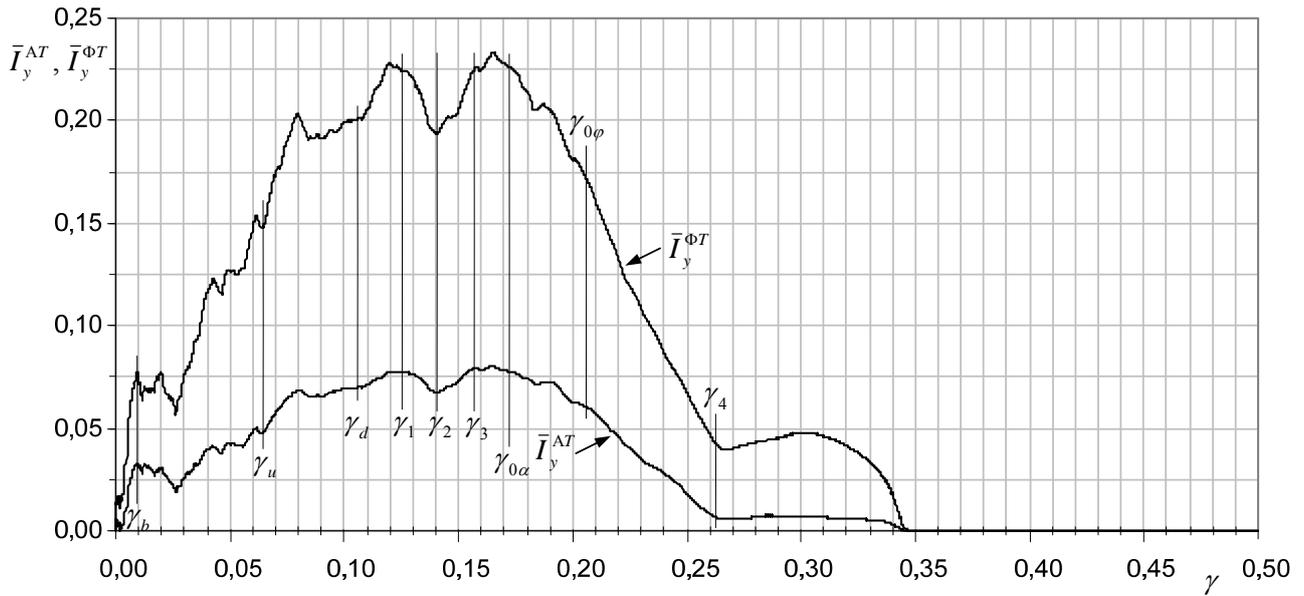

**Fig. 4.** Plot of the intensities $\bar{I}_y^{AT}$ and $\bar{I}_y^{\Phi T}$ versus coupling parameter $\gamma$.
With using the normalized time series (1).

In order to establish the nature of differences between the attractors of process $y(t)$ for various $\gamma$, let us analyze the following quantity:

$$S_y^w(w) = \int_T \delta[w - w_y(t)]dt, \qquad (7)$$

where $\delta$ is the Dirac delta function and $w_y$ is a certain characteristic of the $y(t)$ signal. Figure 5 shows the $S_y^\alpha(\alpha_y^T)$ and $S_y^\varphi(\varphi_{0,y}^T)$ spectra for the $\alpha_y^T$ and $\varphi_{0,y}^T$ components. An analysis of these patterns (with comparison of the left and right peaks to the central one) reveals a significant difference between the structures of attractors of the driven system for the coupling parameters $\gamma_3$ and $\gamma_4$.

The spectrum of shows that the $y(t)$ trajectory at $\gamma = \gamma_3$ contains a large number of steep fronts, both negative ($p_{1\alpha}$, $p_{2\alpha}$) and positive ($p_{3\alpha}$, $p_{4\alpha}$). On the other hand, the spectrum of shows the presence of a large number of sharp pulses with maximum curvature over the entire period $T$, with the orientation of pulses both upward ($p_{1\varphi}$, $p_{2\varphi}$) and downward ($p_{3\varphi}$). Note that the spectra of and at $\gamma = \gamma_2$ qualitatively correspond to those at $\gamma = \gamma_3$, while the spectral lines $p_{2\alpha}$ and $p_{3\alpha}$ decrease to ~1.8 and ~3.1, respectively, the spectral lines $p_{1\varphi}$ and $p_{3\varphi}$ possess virtually identical intensities ~8.5, and spectral line $p_{2\varphi}$ vanishes. These data fully confirm the assumption [4] that the coupling parameter $\gamma_2 = 0.14$ corresponds to an unstable quasi-periodic motion in the driven subsystem, in which state the trajectory $y(t)$ occurs for a significant period of time.

A significant decrease in $I_y^{\circ T}$ at $\gamma = \gamma_4$ can be explained by the appearance of smooth oscillations on trajectory $y(t)$ in the driven subsystem, the total duration of which is rather significant. Indeed, the $S_y^\alpha$ spectrum (Fig. 5a) displays $p_{6\alpha}$ line with weakly positive fronts and a surrounding pedestal (containing weakly negative fronts). The $S_y^\varphi$ spectrum (Fig. 5b) reveals $p_{4\varphi}$ lines with smoothened tops of oscillation pulses oriented with their tops upward and a surrounding pedestal containing smoothened tops of oscillation pulses oriented with their tops downward. In addition, the $S_y^\varphi$ spectrum on the whole is indicative of a more even character of evolution of the $y(t)$ trajectory, with degenerate block of lines adjacent to $p_{1\varphi}$ and $p_{2\varphi}$ at a significant decrease in the intensity of $p_{3\varphi}$ line.

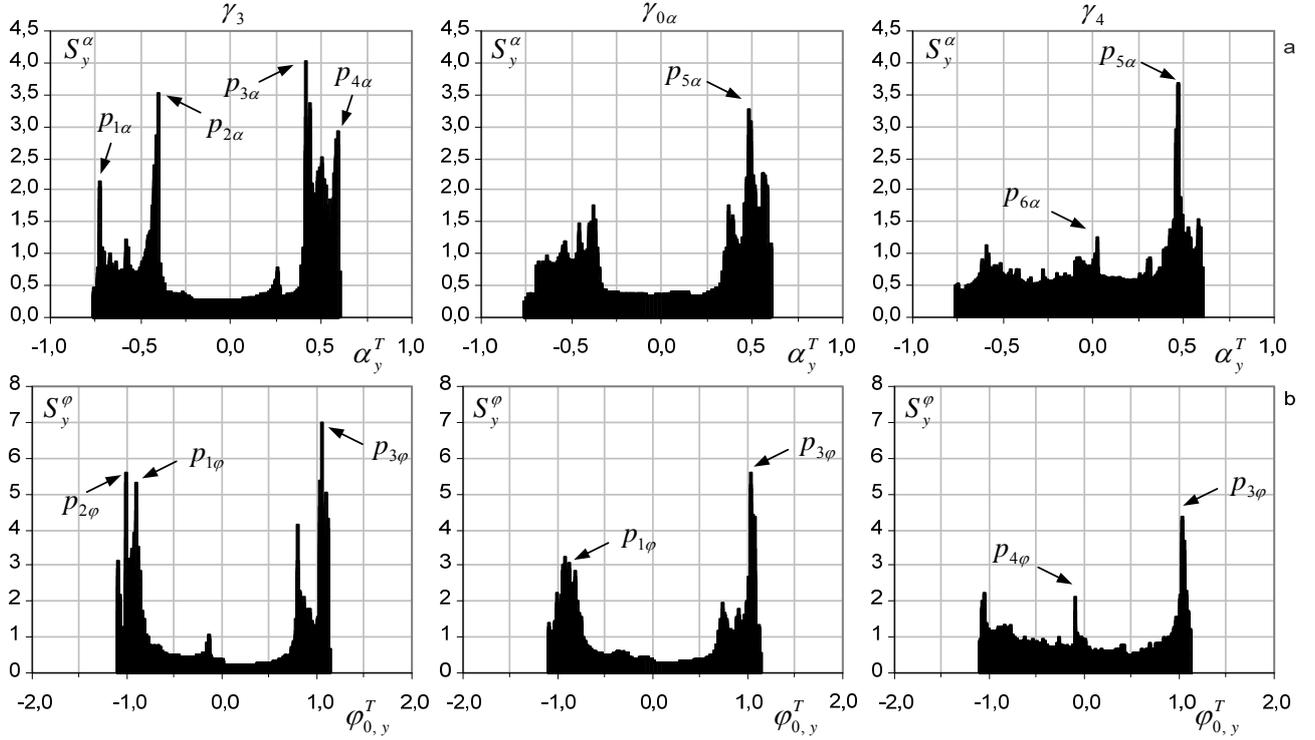

**Fig. 5.** Plots of (a) $S_y^\alpha(\alpha_y^T)$ and (b) $S_y^\varphi(\varphi_{0,y}^T)$ for some specific values of coupling parameter $\gamma$.

It should be noted that an additional information concerning the structure of attractors of trajectory $y(t)$ can be obtained from an analysis of the double spectra $S_y^\alpha(\alpha_y^T, \varphi_{0,y}^T)$ and $S_y^\varphi(\alpha_y^T, \varphi_{0,y}^T)$.

In conclusion, a new approach to the quantitative analysis of the degree and parameters of synchronization of the chaotic oscillations in two coupled oscillators is proposed, which makes it possible to reveal changes in the structure of attractors in the driven subsystem. The proposed method was successfully tested on a model system of two unidirectionally coupled logistic maps in the course of their escape from the state of complete (generalized) synchronization. It is shown that the method is robust with respect to both the presence of a low-intensity noise and a nonlinear distortion of the analyzed signal. The proposed method reveals changes in the structure of attractor in the driven system depending on the coupling parameter and allows the nature of these changes to be analyzed. It is planned to expand the proposed approach to investigations of multidimensional systems and different types of synchronization.

*Translated by P. Pozdeev*

**Andrey V. Makarenko** – was born in 1977, since 2002 – Ph. D. of Cybernetics. Founder and leader Research & Development group "Constructive Cybernetics". Author and coauthor of more than 50 scientific articles and reports. Associate Member IEEE (IEEE Systems, Man, and Cybernetics Society Membership). Research interests: analysis of the structure dynamic processes, predictability; detection, classification and diagnosis is not fully observed objects (patterns); synchronization in nonlinear and chaotic systems; system analysis and modeling of economic, financial, social and bio-physical systems and processes; system approach to development, testing and diagnostics of complex information-management systems.